\documentclass[12pt,reqno]{amsart}
\usepackage{amsmath,amssymb}

\usepackage{enumerate}

\def\dd{\mathrm d}

\def\CC{{\mathbb C}}
\def\RR{{\mathbb R}}

\def\EE{\Bbb E}
\def\CC{\Bbb C}
\def\RR{\Bbb R}

\def \bee{{\mathbf e}}

\def\Ker{{\mathrm{Ker}}}

\def\book#1{{\textit{#1}}, }
\def\paper#1{\textit{#1}, }
\def\jour#1{\rm{#1}, }
\def\yr#1{({\rm{#1}) }}
\def\vol#1{\textbf{#1}}
\def\pages#1{\rm{#1}}
\def\page#1{\rm{#1}}

\def\publaddr#1{\rm{#1}, }
\def\publ#1{\rm{#1}, }
\def\by#1{{\rm{#1}, }}
\def\Bysame{{\bysame,\ }}

\def\Cite#1{[#1]}

\begin{document}

\title{
On an Algebraic Essential of Submanifold Quantum Mechanics }

\author{Shigeki Matsutani}

\thanks{8-21-1 Higashi-Linkan Sagamihara 228-0811 JAPAN\\
\ \ e-mail: RXB01142\@nifty.com}

\vskip 0.5  cm

\maketitle

\begin{abstract}
The submanifold quantum mechanics was opened by Jensen and Koppe
(Ann. Phys. {\bf 63} (1971) 586-591)  and has been studied
for these three decades.
This article gives its more algebraic definition
and show what is the essential of the submanifold quantum mechanics
from an algebraic viewpoint.

\end{abstract}

\bigskip

%\subheading{PACS numbers}:
{\centerline{\textbf{ MCS Codes:}  34L40, 35Q40, 81T20, 32C25}}
%*32C38 Sheaves of differential operators and their modules
%*34L40 Particular operators (Dirac, one-dimensional Schr{\"o}dinger, etc.)
%*35Q40 Equations from quantum mechanics
% 32C25 Analytic subsets and submanifolds
% 32G10 Deformations of submanifolds and subspaces
%*53B25 Local submanifolds [See also 53C40]
%46N50 Applications in quantum physics
% 53C12 Foliations (differential geometric aspects) [See also 57R30, 57R32]
% 53C42 Immersions (minimal, prescribed curvature, tight, etc.)
% [See also 49Q05, 49Q10, 53A10, 57R40, 57R42]
%81T20 Quantum field theory on curved space backgrounds
% 58D10 Spaces of embeddings and immersions

{\centerline{\textbf{ Key Words:}
Laplacian, Schr{\"o}dinger operator, Submanifold}}

\vskip 0.5 cm

The submanifold quantum mechanics we have called was given by
Jensen  and Koppe \cite{JK}
and de Casta \cite{dC} and has been developed by
 Duclos, Exner, Krej{\u c}i{\u r}\'ik, {\u S}eba and
 {\u S}{\u t}ov\'i{\u c}ek \cite{DEK, DES, ES, K},
Ikegami, Nagaoka,  Takagi and Tanzawa \cite{IN, INTT, TT},
Clark and Bracken \cite{C, CB},
Goldstone and Jaffe \cite{GJ}, Bergress and Jensen \cite{BJ},
 Encinosa and Etemadi \cite{EE}, Mladenov \cite{Ml},
Suzuki, Tsuru and this author
\cite{ M1, M2, M3, M4, M5, M6, M7, M9, MT1, MT2, SM}
 and so on.
In these theories, we obtain differential operators
over a submanifold $S$ in an euclidean space
using a confinement potential and taking squeezing limit
of the potential
under some approximating theories.

However the obtained differential operators do not strongly
depend upon the shape of confinement potentials
or ways to take  squeezing limits.
Further they exhibit geometrical nature of the submanifold.
In fact, the Dirac operators obtained in the scheme
are related to the Frenet-Serret and the
generalized Weierstrass relations
\cite{M2, M3, M4, M5, M6, M7, M8, M9}:
they recover all geometrical data of submanifold.
Thus we believe that it should be obtained beyond an approximation
and defined more algebraically.

In this article, we will give a more algebraic definition
of  the submanifold quantum mechanics, which is free from
any approximation theories, and
show what is the essential
of the submanifold quantum mechanics
from an algebraic point of view.

\vskip 1.0 cm

As shown in later,
in submanifold quantum mechanics,
 a non-unitary transformation
plays a key role, which makes  a not self-adjoint
operator  self-adjoint.
Thus as a preparation,
we should investigate the self-adjoint operator precisely.
However the concept of \lq\lq{adjointness}\rq\rq of an operator
is subtle even in study of algebra of differential operators
as in [Remark 1.2.16 in \cite{Bj}].
Let us consider a differential operator
$\partial/\partial z^\alpha$ defined
 over a $n$-dimensional
differential manifold $M$ with local coordinate $z$.
Let $\partial/\partial z^\alpha$
 act a function from left hand side conventionally.
Provided that  it is equipped with a metric $(g)$ and
a volume form $ g^{1/2} d^n z$,
for smooth wavefunctions $f_1$ and $f_2$ whose support is compact,
we have a natural pairing, a map to the complex number,
\begin{gather}
<f_1| f_2>_g
=\int_M g^{1/2} \dd^n z\ f_1^*(z) f_2(z). \tag{1}
\end{gather}
Using the pairing,
an expectation value of $\partial/\partial z^\alpha$ is
also naturally defined by,
\begin{gather*}
<f_1|\frac{\partial}{\partial z^\alpha} f_2>_g
=\int_M g^{1/2} \dd^n z\ f_1^*(z)\frac{\partial}{\partial z^\alpha}
 f_2(z).
\end{gather*}
The adjoint operator of $\partial/\partial z^\alpha$ is given as
$(\partial/\partial z^\alpha)^*$ $=
-\partial/\partial z^\alpha -1/2 (\partial\log g/\partial z^\alpha)$,
which depends upon the measure.
For another  measure  ${g'}^{1/2} \dd^n z$
such that $g^{1/2} \dd^n z\ f_1^*(z) f_2(z)
={g'}^{1/2} \dd^n z\ (\alpha f_1)^*(z) (\alpha f_2)(z)$
where $\alpha := (g/g')^{1/4}$, we might have another expectation
value,
\begin{gather}
<f_1|\frac{\partial}{\partial z^\alpha} f_2>_{g'}
=
\int_M {g'}^{1/2} \dd^n z\ f_1^*(z)
 \frac{\partial}{\partial z^\alpha}
f_2(z). \tag{2}
\end{gather}
we have a different adjoint operator
$(\partial/\partial z^\alpha)^*$.
The  adjointness has such an ambiguity.

Using the ambiguity, we can introduce the half-density
and then any $\partial/\partial z^\alpha$ can be self-adjoint
by setting $g'=1$ (Theorem 18.1.34 in \cite{H}).
However in general the measure $g^{1/2} \dd^n z$,
{\it e.g.}, Haar measure,
exhibits a geometrical nature of the space.
Thus the measureless expression such as
the half-density does sometimes
have less effective \Cite{N1}.

In the submanifold quantum mechanics \cite{dC, JK},
we partially use the half-density in the
framework of a theory with the Haar measure
 as we will show later.

From quantum mechanical point of view, the first problem
on quantum mechanics over a curved system is to
search a proper metric and a proper measure.
This problem is easily solved in
 quantum mechanics over a curved object in our euclidean space $\EE^3$.
In our euclidean space $\EE^3$, the ordinary Lebesgue measure
is natural because it is the  Haar measure for the translation.
As the quantum mechanics is established in $\EE^3$
and the concept of the adjoint operator plays essential roles
\cite{D},
we will use an induced metric on the curved object from
that in $\EE^3$.

Next we will review  properties
 of a self-adjoint operator precisely.
We deal with function spaces $\Omega$ and its dual $\Omega^*$
with L$^2$-type paring $<,>: \Omega^* \times\Omega \to \CC$
using the Lebesgue measure in $\EE^3$. We consider
 an operator $Q$ whose domain is $\Omega$.
Then if exists, we could define a right-adjoint operator, $Ad(Q)$,
with the domain $ \Omega^*$ by
\begin{gather}
       <f,Q g> = < f Ad(Q), g> , \quad\text{for }
       (f,g)\in \Omega^* \times\Omega.
 \tag{3}
\end{gather}
Assume that there is an isomorphism $\varphi$
between domains $\Omega$  and $  \Omega^*$ as a vector space
\Cite{N2}.
Then triplet
$(\Omega^*\times \Omega, <,>, \varphi)$ becomes
 a Hilbert space $\mathcal H$ by introducing
the inner product $(,):\Omega\times\Omega \to \CC$ be
$(f,g):=<\varphi(f),g>$ after completion in $(,)$.
We assume such completion in this article and thus we write
$(\Omega^*\times \Omega, <,>, \varphi) =\mathcal H$ hereafter.
Then the ordinary adjoint operator $Q^*$,
$(Q^* f, g)=(f,Q g)$,
 is given by
$Q^*f := \varphi^{-1}(\varphi(f) Ad(Q))$.

Suppose that the operator $Q$ is {\it self-adjoint},
{\it i.e.}, the domains  of $Q^*$ and $Q$ coincide
and $Q^*=Q$ over there.
Then we have the following properties:

\begin{enumerate}
\item The kernel of $Q$, $\Ker (Q)$, is isomorphic to
the $\Ker (Ad(Q))$ {\it i.e.},
\begin{gather}
	(\Ker (Q))^* :=
	\varphi(\Ker (Q)) = \Ker( Ad(Q)).
        \tag{4}
\end{gather}
\item
The projection $\pi_Q$ from $\Omega^*\times\Omega$ to
$(\Ker (Q))^* \times \Ker (Q)$  is
commutative with $\varphi$, {\it i.e.},
\begin{gather}
	\varphi \pi_Q|_\Omega=\pi_Q|_{\Omega^*} \varphi , \quad
       (\varphi( \pi_Q|_\Omega f) = \pi_Q|_{\Omega^*}
          \varphi(f) \equiv \varphi(f) Ad( \pi_Q|_\Omega)) .
 \tag{5}
\end{gather}

\item $( (\Ker (Q))^*\times \Ker (Q), <,>, \varphi)$ becomes
 a Hilbert space.

\end{enumerate}

For $\pi_Q$ satisfying (5), we will
say that {\it $\pi_Q$ is consistent with the inner product}.
In fact (5) means that $\varpi_Q:=\pi_Q|_\Omega$ is a projection operator
as $*$-algebra \cite{A}: $\varpi_Q^2=\varpi_Q$ and $\varpi_Q^*=\varpi_Q$
due to the relation
$\varpi_Q^*f \equiv \varphi^{-1}(\varphi(f)Ad(\varpi_Q))=\varpi_Q f$.

As we finished the review of the properties of self-adjoint
operators,
  let us give a physical setting on the submanifold quantum
mechanics.
Though we can do more general, we will investigate
only a case of
a surface in three dimensional euclidean space $\EE^3$.
For a smooth surface $S$ embedded in the
euclidean space
$\EE^3$ with the induced metric of $S$ from that in
$\EE^3$,
we consider a Schr{\"o}dinger equation
over a tubular neighborhood $T_S$ of $S$,
$\pi_{T_S}:T_S \to S$, with
the L$^2$-type Hilbert space $\mathcal
H=(\Omega^*\times \Omega, <,>_{g_{T_S}}, \varphi)$,
\begin{gather}
	-\Delta \psi = E\psi \quad \text{over}\quad
              T_S. \tag{6}
\end{gather}
Here $\Omega^*$ and $\Omega$ consist of smooth
compact support functions over $T_S$ and $\Delta$ is
the Laplacian in $\EE^3$.
In the ordinary methods \cite{dC, JK},
we add a potential to the left hand side in (6),
which confines a particle in the
tubular neighborhood $T_S$.
The potential makes the support of the wavefunctions
in $T_S$. By taking a squeezing limit
of the potential, we decompose the system
to the normal and the
tangential modes, suppress the normal mode,
 and obtain the submanifold
Schr{\"o}dinger equation along the surface $S$.
Instead of the scheme, we will give another
definition of the Schr{\"o}dinger operator by (16).

Before giving a novel definition,
 we will give a geometrical preliminary.
Let $S$ be locally expressed by a coordinate system
$(s^1,s^2)$
and $q$ be a normal coordinate of $T_S$ whose absolute
value is the distance from the surface $S$;
$dq$ is an infinitesimal length in $\EE^3$
and belongs to kernel of $\pi_{T_S*}$ and
$dq( \partial_{\alpha})=0$ $(\alpha=1, 2,
\partial_\alpha := \partial/\partial s^\alpha)$.
A point in $T_S$  expressed by the
affine coordinate ${\mathbf x}:=(x^1,x^2,x^3)$ in $\EE^3$
can be uniquely represented by
\begin{gather}
	{\mathbf x}=\pi_{T_S}{\mathbf x}+q {\mathbf e_3},
\tag{7}
\end{gather}
where ${\mathbf e_3}$ is the normal unit vector at $S$.
The moving frame
  $E^i_{\ \alpha}=\partial_\mu x^i$, $(\mu =1,2,3: i=1,2,3)$
  is written by,
\begin{gather}
    E^i_{\ \alpha} = e^i_{\ \alpha} +
            q^3\gamma^\beta_{\
            3\alpha} e^i_{\ \beta},
            \quad  E^i_{\ 3} = e^i_{\ 3},
          \tag{8}
\end{gather}
where $\alpha, \beta=1,2$,
$e^i_{\ \alpha}:=\partial_\alpha (\pi_{T_S}{x^i})$,
 and
\begin{gather}
    \partial_\alpha \bee_3 =
    \gamma^\beta_{\ 3\alpha}
  \bee_{\beta}.\tag{9}
\end{gather}

Thus the induced metric, $g_{T_S \mu\nu}$
$:=\delta_{i j}E^i_\mu E^j_\nu$, ($\mu, \nu = 1, 2, 3$),
from that in the euclidean space $\EE^3$
is given as
\begin{gather*}
 g_{T_S}=\begin{pmatrix} g_{ {S_q}} & 0 \\ 0 & 1
\end{pmatrix},
\end{gather*}
\begin{gather}
g_{ {S_q} \alpha\beta}
  = g_{S \alpha\beta}+
    [\gamma_{\ 3\alpha}^\gamma g_{S\gamma\beta}+
    g_{S\alpha\gamma}\gamma_{\ 3\beta}^\gamma]q
    +[\gamma_{\ 3\alpha}^\delta g_{S\delta\gamma}
     \gamma_{\ 3\beta}^\gamma]q^2 ,
     \tag{10}
\end{gather}
where $g_{S \alpha\beta}:=\delta_{i j}e^i_\alpha e^j_\beta$.
The determinant of the metric is expressed as,
\begin{gather}
\det g_{T_S}=\rho \det g_S ,\quad
\rho = (1 + \mathrm{tr}(\gamma_{\ 3\beta}^\alpha)q
          + \mathrm{det}(\gamma_{\ 3\beta}^\alpha)q^2),
        \tag{11}
\end{gather}
and thus the pairing $<,>_{g_{T_S}}$ is expressed by
\begin{gather}
	<\psi_1,\psi_2>_{g_{T_S}} = \int (\det g_{S})^{1/2} \rho^{1/2}
       \dd^2 s \dd q \ \psi_1^* \psi_2.
      \tag{12}
\end{gather}

As the titles in \cite{dC, JK}, we wish to
establish the quantum mechanics
of a particle restricted at $S \subset \EE^3$.
The restriction of the particle into the surface requires
that momentum and position of the particle for the normal
direction vanish.
In order to realize  the vanishing momentum, we will
consider the momentum of the normal direction.
We have
the canonical commutation relation for the normal direction
\Cite{N3},
$[\sqrt{-1}\partial_q, q]\equiv
\sqrt{-1}\partial_q q - q \sqrt{-1}\partial_q=\sqrt{-1}$
and wish to consider kernel of $p_q:=\sqrt{-1}\partial_q$.
However $p_q$ is not self-adjoint in general
due to the existence of $\rho$ in (12)
and thus a projection to
kernel of $\partial_q$ must not be consistent with
the inner product in the above sense.

Thus we deform the Hilbert space so that
$p_q$ becomes a self-adjoint operator
by using the half-density theory
(Theorem 18.1.34 in \cite{H}).
As we will show (13) and (14), there exist
a Hilbert space $\mathcal H'
\equiv (\tilde \Omega^* \times\tilde  \Omega,
<,>_{g_S}, \tilde\varphi)$ and
{\it self-adjointization}:
$\eta_{\mathrm{sa}} : \mathcal H \to \mathcal H'$
satisfying the following properties.
\begin{enumerate}

\item There exists an isomorphism
$\Omega^* \times \Omega \to \tilde \Omega^* \times\tilde  \Omega$
as a vector space. We denote it by the same
$\eta_{\mathrm{sa}}$.

\item By defining a pairing
$<\circ,\times>_{g_S}
:=<\eta_{\mathrm{sa}}\circ,\eta_{\mathrm{sa}}\times>_{g_{T_S}}$,
we set $\tilde \varphi:= \eta_{\mathrm{sa}} \varphi \eta_{\mathrm{sa}}^{-1}$.

\item
An operator $Q$ of $\mathcal H$ is transformed into
$\eta_{\mathrm{sa}}(Q)=\eta_{\mathrm{sa}} Q \eta_{\mathrm{sa}}^{-1}$.

\item $p_q$ itself (not $\eta_{\mathrm{sa}}(p_q)$)
becomes a self-adjoint operator in $\mathcal H'$.

\end{enumerate}
Of course, the self-adjointization is not a unitary operation
and due to the operation, the inner product changes from
$<,>_{g_{T_S}}$
to $<,>_{g_S}$.

Using the dependence of the adjoint operator upon the measure
as mentioned above,
the self-adjointization $\eta_{\mathrm{sa}}$ is realized as follows:
\begin{gather}
	\eta_{\mathrm{sa}}(\psi_1^*) = \rho^{-1/4} \psi_1^*, \quad
	\eta_{\mathrm{sa}}(\psi_2) = \rho^{-1/4} \psi_2, \quad
 	\eta_{\mathrm{sa}}(\Delta) =\rho^{-1/4} \Delta \rho^{1/4},
       \tag{13}
\end{gather}
\begin{gather}
	<\psi_1,\psi_2>_{g_S} = \int (\det g_{S})^{1/2}
	                  \dd^2 s \dd q \ \psi_1^* \psi_2.
        \tag{14}
\end{gather}
In $T_S$, $\rho$ does not vanishes and (13) give isomorphisms.
Since in the measure of (14),  $q$-dependence disappears,
$p_q:=\sqrt{-1}\partial_q$ itself becomes self-adjoint in $\mathcal H'$
\Cite{N4}.
Hence the projection,
\begin{gather}
	\pi_{p_q}: \tilde \Omega^* \times \tilde \Omega \to
          (\Ker( Ad(p_q)))^*\times \Ker (p_q), \tag{15}
\end{gather}
is consistent with the inner product $(,)$.
Due to self-adjointness, $\tilde\varphi: \Ker (p_q) \to
\Ker (Ad(p_q))$ is isomorphic each other \Cite{N5}.
In other words,
${\mathcal H}_{p_q}:=
( (\Ker (Ad(p_q)))\times \Ker (p_q), (,),\tilde\varphi)$
becomes a small quantum mechanical system.

We should note that as a Goldstone mode is
sometimes given as a zero mode of differential
operator exhibiting a symmetry \cite{Bu}, $\psi \in \Ker (p_q)$ behaves
like a Goldstone mode for the normal translation mode.
In fact, $p_q$ mode does not
contribute in $\eta_{\mathrm{sa}}(\Delta)$
over the small Hilbert space ${\mathcal H}_{p_q}$.
The projection to ${\mathcal H}_{p_q}$ means vanishing momentum of
the normal direction and
kills a normal translation freedom.
Hence we can choose a position $q$
as an ordinary symmetry breaking \Cite{N6}.
After choosing it as $q=0$, the Laplacian in (6) becomes
\begin{gather}
	\Delta_{S \hookrightarrow \EE^n}
         :=\eta_{\mathrm{sa}}(\Delta)|_{\Ker( p_q)}|_{q=0},
        \tag{16}
\end{gather}
as an operator in ${\mathcal H}_{p_q}|_{q=0}$ \Cite{N7, N8}.
By letting
$K$ and $H$ denote the Gauss and mean curvatures
 of $S \subset \EE^3$,
we obtain the well-known operator \cite{dC, JK},
\begin{gather}
	\Delta_{S \hookrightarrow \EE^3}
         =
	\Delta_{S} +  H^2-K, \tag{17}
\end{gather}
and the submanifold Schr{\"o}dinger equation,
\begin{gather}
	-\Delta_{S \hookrightarrow \EE^3} \psi
= E\psi \quad \text{over } S, \tag{18}
\end{gather}
over the Hilbert space ${\mathcal H}_{p_q}|_{q=0}$.
Here $\Delta_{S}$ is the Beltrami-Laplace operator on $S$
which exhibits the intrinsic properties of the surface $S$,
whereas the second and the third terms in (17)
represent the extrinsic properties
of $S\subset \EE^3$.

Here we emphasize that the definition (16) is very algebraic.
Particularly
in this construction,  we did not use
any approximation theories nor limit-theorems.
Physically speaking, the above requirement
of vanishing momentum and position might be
 contradict with
the uncertainly principle. However the vanishing
normal momentum naturally leads a symmetry breaking.
Thus if we regard the normal direction as an inner space,
we believe that the above requirement is natural.
In fact
our construction is consistent with that
in \cite{INTT}, in which (18) was obtained by
means of the Dirac constraint quantization scheme
under the constraint condition of vanishing momentum.

Further our study reveals why we need deform the Hilbert space
in the construction of the submanifold quantum mechanics
\cite{dC, JK} \Cite{N9}.
Hence we show the algebraic essential
of the submanifold quantum mechanics.

By means of the construction, we can give a
more algebraic representation of generalized Weierstrass relation
in terms of the submanifold Dirac operators \cite{M5, M6, M7, M9},
which is closely related to the extrinsic Polyakov string
\cite{M8}.

\bigskip
\bigskip

\begin{centerline}{\bf Notes}
\end{centerline}

\begin{itemize}
\item[{[N1]}]
For example, in the theory of the ordinary second order
differential equation related to orthonormal polynomial
functions, the concept of the half-density implicitly
appears (p.424 in \cite{Arf}) but gives only
very rough estimations, such as the asymptotic expressions
of the functions.
On the other hand, their expressions
with the proper measure as the orthonormal polynomials
give more precise information.
In other words, measureless expression is not sometimes
effective to concrete problems well.

\item[{[N2]}]
In general, the dual space of $\Omega$ is greater than
$\Omega$ itself, thus we regard  $\Omega^*$ as the image
of $\varphi$. In this construction,  the
Dirac $\delta$ functions are elements in the complement
of the image of $\varphi$ in the dual space.

\item[{[N3]}]
We note that the canonical commutation relation
has an ambiguity.
For a function $h(q)$ of $q$, $[h(q),q]=0$
and thus we have $[\sqrt{-1}\partial_q+h(q), q]=\sqrt{-1}$.
Using the ambiguity, we can optimize $h$ so that
$\sqrt{-1}\partial_q+h(q)$ is self-adjoint. However
in this article, we give another equivalent scheme called
self-adjointization.

\item[{[N4]}]
Here we should notice that
the proper treatment of self-adjoint operator and a prototype
of self-adjointization are implicitly written in the study
of hydrogen atom in the text book of Dirac \cite{D}.
In the book, instead of $p_q$, the formally self-adjoint
operator $\eta_{\mathrm{sa}}^{-1}(p_q)$ is treated.
$\eta_{\mathrm{sa}}^{-1}(p_q)$
determines an addition term $h(q)$
to $\sqrt{-1}\partial_q$ \Cite{N3}
so that it is self-adjoint. Corresponding to the property (4)
of $\eta_{\mathrm{sa}}$, the transformation from
$\sqrt{-1}\partial_q$ to $\eta_{\mathrm{sa}}^{-1}(p_q)$
is not unitary.

\item[{[N5]}]
{
Precisely we should write the map
$\tilde\varphi|_{\Ker (p_q)}$ instead of $\tilde\varphi$.}

\item[{[N6]}]
{Precisely speaking,
since $\eta_{\mathrm{sa}}(\Delta)|_{\Ker( p_q)}$ has $q$-dependence
and the energy weakly depends upon $q$,
it slightly differs from the ordinary symmetry breaking.
However as the dependence is not strong, it can be justified.
If we choose $q=q_0$ for all points in $S$,  we have
submanifold Schr{\"o}dinger equation over a surface $S_{q_0}$
given by $q=q_0$ instead of the surface $S$ for $q=0$.}

\item[{[N7]}]
The first restriction $|_{\Ker{p_q}}$ means the restriction
of domain as an operator and the second one
$|_{q=0}$ should be regarded as an restriction in
the meaning of presheaf theory.

\item[{[N8]}]
Since (16) can be written by
\begin{gather*}
	\Delta_{S \hookrightarrow {\mathbf E}^n}
         :=(\rho^{1/4}\Delta \rho^{-1/4})|_{\partial_q = 0, q=0},
\end{gather*}
it might be expected that the definition
 should be expressed in differential ring
theory \cite{Bj}.

\item[{[N9]}]
If we did not deform the Hilbert space
using $\eta_{\mathrm{sa}}$, $\varpi_{p_q}$ cannot
becomes the projection operator in the sense of
$*$-algebra \cite{A}.

\end{itemize}

\end{document}